\documentclass[]{aa}

\usepackage{graphicx}

\newcommand{\ten}[1] {10$^{#1}$}
\newcommand{\WmAarcsec}{W m$^{-2}$ \AA$^{-1}$ arcsec$^{-2}$}
\newcommand{\Wmarcsec}{W m$^{-2}$ arcsec$^{-2}$}
\newcommand{\kms}{km~s$^{-1}$}
\newcommand{\etal}{et~al.}
\newcommand{\Ha}{\hbox{H$\alpha$}}

\newcommand{\NII}{\hbox{[N\,{\sc ii}]}}

\newcommand{\NIIwb}{\hbox{[N\,{\sc ii}]$\lambda $6583}}
\newcommand{\NIIww}{\hbox{[N\,{\sc ii}]$\lambda\lambda $6548,6583}}

\newcommand{\SIIww}{\hbox{[S\,{\sc ii}]$\lambda\lambda $6717,6731}}

\begin{document}

\thesaurus{03(11.09.1 M104; 11.05.2; 11.11.1; 11.14.1; 11.16.1; 11.19.5)}

\title{The Sombrero galaxy
\thanks{Based on observations taken with the Canada-France-Hawaii
Telescope, operated by the National Research Council of Canada,
the Centre National de la Recherche Scientifique of France, and the University of
Hawaii, and observations with the NASA/ESA {\it Hubble Space telescope} obtained at
the Space Telescope European Coordinating Facility, jointly operated by ESA
and the European Southern Observatory.}}

\subtitle{III. Ionised gas and dust in the central 200~pc: a nuclear bar?}

\author{E. Emsellem \inst{1} \and P. Ferruit \inst{1}}

\offprints{E. Emsellem (email: emsellem@obs.univ-lyon1.fr)}

\institute{Centre de Recherche Astronomique de Lyon, 9 av. Charles Andr\'e,
	69561 Saint-Genis Laval Cedex, France}

\date{Submitted to A\&A Main Journal}

\maketitle
\markboth{Emsellem \& Ferruit: Ionised gas and dust in the central 200~pc of M~104}{}

\begin{abstract}
   We present the results of new 3D TIGER spectroscopic observations and archived
   HST/WFPC2 and NICMOS images of the central region of M\,104. The \NII+\Ha\ images
   reveal the presence of a nuclear spiral structure, and the gaseous kinematics in the
   central arcsecond shows evidence for kinematical decoupling of the central peak.
   A straight nuclear dust lane, with a weak symmetric counterpart, is seen in the
   $V-I$ and $V-H$ colour maps. These results hint for the presence of a strong
   nuclear bar, that would be located inside the inner Linblad resonance of the
   large-scale bar discussed by Emsellem (\cite{Ems95}).

\end{abstract}

\section{Introduction}
   This is the third of a series of papers dealing with optical observations of the
   Sombrero galaxy (M~104, NGC~4594). We report here the results of 3D spectroscopy of
   the nuclear region of M\,104 with the TIGER instrument, in the 6750/460~\AA\
   spectral domain, which includes the \NIIww, \Ha\
   and \SIIww\ emission lines. This 3D dataset is used in combination with archived
   images acquired with the Wide Field and Planetary Camera 2 (WFPC2) and the Near
   Infrared Camera and Multi-Object Spectrometer (NICMOS) on board the Hubble Space
   Telescope (HST), to study the nuclear regions of this galaxy. This paper is
   organised as follows: Sect.~\ref{SecObs} presents the observations and the data
   reduction; Sect.~\ref{SecRes} shows the results; and Sect.~\ref{SecDis} contains a
   brief discussion of the nature of the observed nuclear structures.
   
   Throughout this paper, we will use a distance to M~104 of 8.8~Mpc (Ciardullo \etal\
   \cite{Ciar93}), yielding an intrinsic scale of 
	$\sim$43~pc.arcsec$^{-1}$. We will also use a systemic (heliocentric) velocity $V_s$ of
   1080~\kms, infered from the two-dimensional stellar velocity field (Emsellem \etal\
   \cite{Ems96}, hereafter Paper~2), and a value of  84\degr\ ($\pm 2$\degr) for the
   inclination of the galaxy (Paper~2).

      \begin{figure*}
	 \includegraphics[width=\textwidth,clip=true,draft=false]{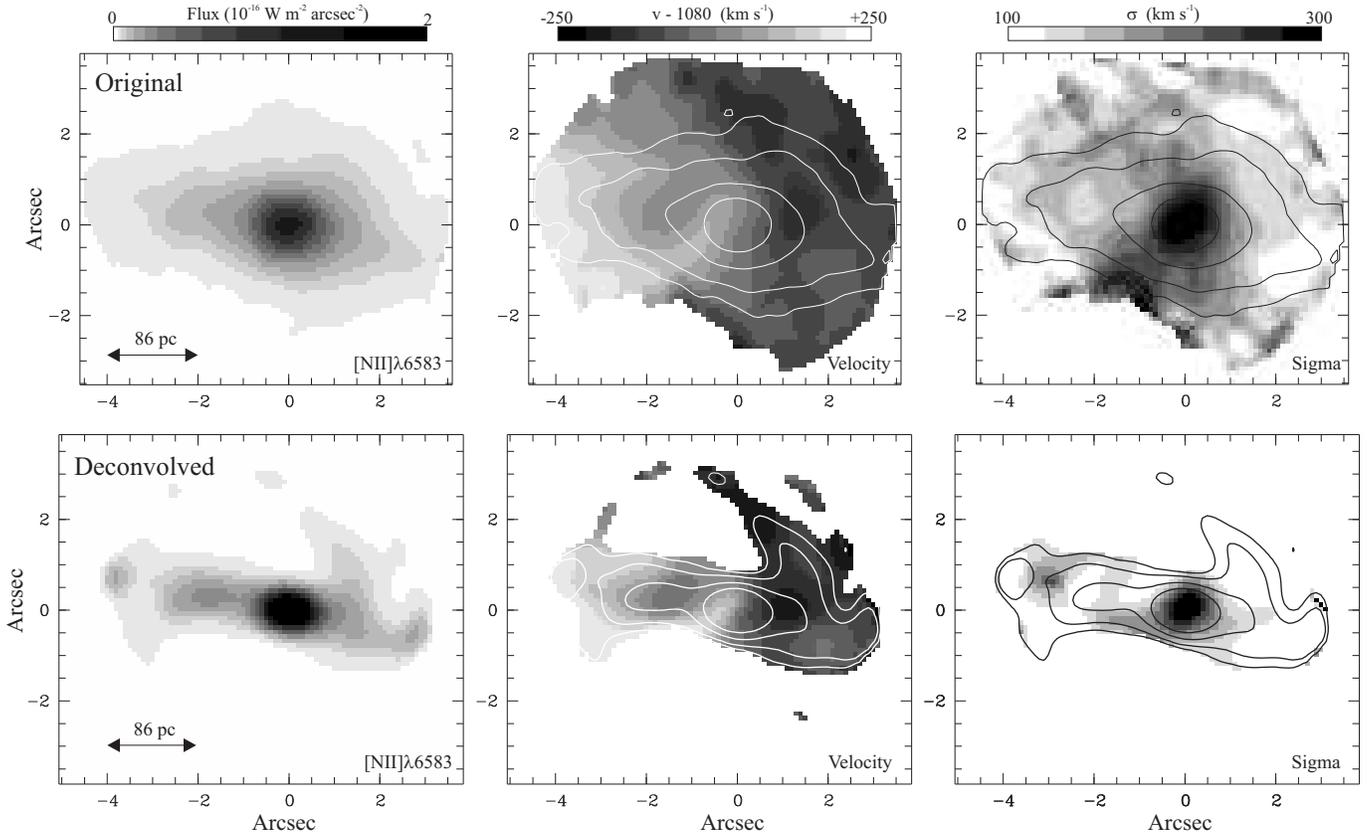}
	 \caption{{\bf Top row:} \NIIwb\ line flux (left), heliocentric 
	 centroid  velocity $V$ (middle, minus $V_s =  1080$~\kms) and dispersion 
	 $\sigma$ (right, corrected from the instrumental dispersion) maps, as
	 derived from a single component Gaussian fitting of the emission lines in
	 each spectrum of the original (i.e. not deconvolved) TIGER data cube.
	 Black or white contours correspond to selected isophotes (2.5, 5, 15, and 
	 45 $\times$ \ten{-18} \Wmarcsec) of the \NIIwb\ line flux map. 
	 Spectra with a \NIIwb\ line peak intensity $<$ 0.9 $\times$~\ten{-19}
	 \WmAarcsec\ (i.e. $\sim 3\sigma_{\mbox{\tt noise}}$) 
	 have been discarded in the $V$ and $\sigma$
	 maps. {\bf Bottom row:} Same as top row, but derived from the deconvolved data
	 cube (Richardson-Lucy deconvolution, 40 iterations, see
	 Sect.~\protect\ref{SecObsTig}). Spectra with a fitted \NIIwb\ line peak
	 intensity $<$ 0.24 $\times$~\ten{-19} \WmAarcsec\ (i.e. $\sim 3\sigma_{\mbox{\tt noise}}$) 
	 have been discarded in the $V$ and $\sigma$ maps. North is up, East left.}
	 \label{FigNII}
      \end{figure*}
\section{Observations and data reduction}
   \label{SecObs}
   \subsection{HST/WFPC2 and NICMOS imaging}
      \label{SecObsHST}
      From the Space Telescope/European Coordinating Facility (ST/ECF) archive at the
      European Southern Observatory, we have retrieved WFPC2 and NICMOS images of
      M\,104 in four bands: F547M, F658N, F814W, and F160W. Their total integration
      times were 1340, 1120, 1600, and 128 seconds, respectively. The data reduction
      was performed using the ST/ECF pipeline, with the most recent calibration data.
      
      The images were flux calibrated, rotated to the cardinal orientation (north up,
      east to the left), and corrected for geometric distorsion. An emission-line
      image (\NII+\Ha) has been constructed by subtracting the F814W image from the
      on-band F658N image. Colour maps, similar to $V-I$ and $V-H$, were also
      constructed, using the F547M, F814W and F160W  images\footnote{The $V-I$
      map has already been presented by Pogge \etal\ (\cite{Pogge99}) and Kormendy
      \etal\ (\cite{Korm96}). Note that in Kormendy \etal\
      (\cite{Korm96}), the FOS apertures in their Fig.~1 are on the wrong side (east
      and west are reversed).}.
      
   \subsection{TIGER spectrography}
      \label{SecObsTig}
      We obtained a total of 1.5 hour integration on the centre of M\,104 using the
      TIGER 3D spectrograph (Bacon et al. \cite{Bacon95}) at the Canada France Hawaii
      telescope, in April 1996. The spectral domain (6750/460~\AA) includes the
      \NIIww, \Ha\ and \SIIww\ emission lines. The spectral sampling was 1.5~\AA\ per
      pixel, with a spectral resolution of $\sim$3.5~\AA\ FWHM. The data reduction
      was performed using a dedicated software (Rousset \cite{Rousset92}). The spatial
      sampling was $0\farcs39$ per lens, with a (seeing limited) spatial resolution of
      $0\farcs95$. Subtraction of the stellar continuum was achieved via a library of
      stellar and galactic spectra using a procedure which will be detailed in a
      forthcoming paper (Emsellem \etal\ in preparation).
      
      To improve the spatial resolution of this data set, we have used a
      Richardson-Lucy algorithm (Richardson 1972; Lucy 1974) to deconvolve each
      velocity slice of our continuum subtracted data cube. The point-spread function
      used in this deconvolution was obtained by comparing the emission-line image
      derived from the TIGER data cube and the HST emission-line image. We limited the
      number of iterations to 40 and the spatial resolution after deconvolution is
      $\sim 0\farcs5$, with a new spatial sampling of $0\farcs2$ per pixel.
      
      The emission lines present in our spectra have been fitted using the FIT/SPEC
      software (Rousset \cite{Rousset92}) to reconstruct maps of the ionised gas
      distribution and kinematics. Except when explicitely mentioned, we only present
      results obtained by fitting a single Gaussian profile for each individual
      emission line, with all lines constrained to share the same velocity shift and
      width. Excellent fits were obtained, even for the nuclear emission-line
      profiles, despite their extended wings (e.g. see Fig.~3 in Kormendy
      \etal\ \cite{Korm96}, hereafter K+96).

\section{Results}
   \label{SecRes}
   \subsection{Ionised gas distribution and kinematics}
      
      In Fig.~\ref{FigNII}, we present maps of the \NIIwb\ line intensity, centroid
      velocity and velocity dispersion (the latters being common to all fitted lines).
      The centre of the field has been defined as the location of the maximum line
      intensity, which coincides for all observed lines. The ionised gas
      distribution derived from our deconvolved data cube agrees with that obtained in
      the HST \NII+\Ha\
      image (Pogge \etal\
      \cite{Pogge99}), and exhibits three prominent features: a bright core, two
      roughly symmetric spiral-like structures, and a curved extension northwest of
      the nucleus.
         
      \begin{figure}
	 \includegraphics[width=\columnwidth,clip=true,draft=false]{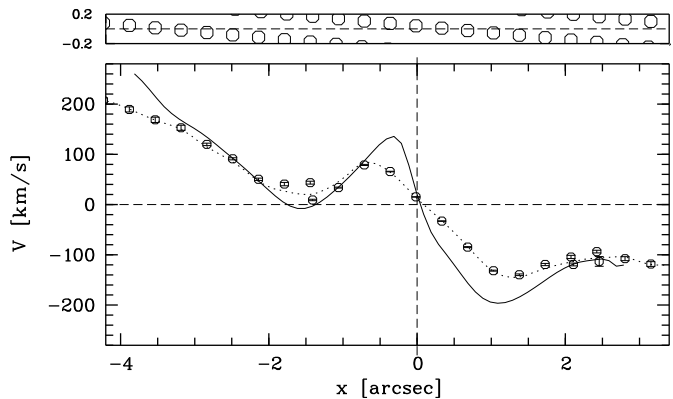}
	 \caption{Plot of the (heliocentric) ionised gas velocity (minus $V_s = 
	 1080$~\kms) along the major axis of the galaxy (east is left). Velocities
	 derived from the spectra within $0\farcs2$ of the major axis (original
	 dataset) are represented by circles, with error bars overimposed. The
	 upper-panel shows the location of these spectra on the sky. We also
	 display the interpolated velocity profiles, as derived from the original
	 (dotted line) and deconvolved (solid line) data cubes.}
	 \label{FigProf}
    \end{figure}
      
      The spiral structures harbor velocities significantly lower than the
      circular velocities predicted by the multi-Gaussian expansion model of M~104
      (Paper~2). Similar behavior has been noticed for gas between 10 and 50\arcsec\
      from the nucleus (Rubin \etal\ \cite{Rubin85}), as well as in nuclear FOS
      spectra (K+96). A cut of the velocity map along the major axis of the galaxy
      (Fig.~\ref{FigProf}), shows the presence of a strong velocity gradient close to
      the nucleus, with extrema of $+130$ and $-200$~\kms, located $0\farcs4$ (18~pc)
      east and $1\farcs1$ (48~pc) west of the centre, respectively. These extrema are
      followed by an abrupt decrease of the velocity modulus, which reaches minima
      $1\farcs4$ (60~pc) east and $2\farcs3$ (100~pc) west of the nucleus, before
      increasing again. The kinematics of the ionised gas  inside $\sim$1\arcsec\
      (43~pc) is thus decoupled from that of the gas in the spiral arms. This
      decoupling is also present in the datacube before deconvolution. After
      deconvolution, the velocity gradient in the central arcsecond is larger on the
      eastern side than on the western side.
      
      The velocity dispersion is centrally peaked, reaching a value of 380~\kms\
      in the deconvolved data. From their high spatial resolution FOS data, Nicholson
      et al. (\cite{Nich98}) obtained larger central velocity dispersions, of 540 and
      390~\kms\ for \NII\ and \Ha, respectively. However, this discrepancy can easily
      be accounted for by the difference in spatial resolution and the fact that we
      did not attempt to model the \NII\ and \Ha\ lines separately. Note also that
      the central velocity dispersion for the gas in our deconvolved data is similar
      to the stellar one from K+96 (FOS data).

   \subsection{Colour map and dust distribution}
      \label{SecDust}
      \begin{figure}
	 \includegraphics[width=\columnwidth,clip=true,draft=false]{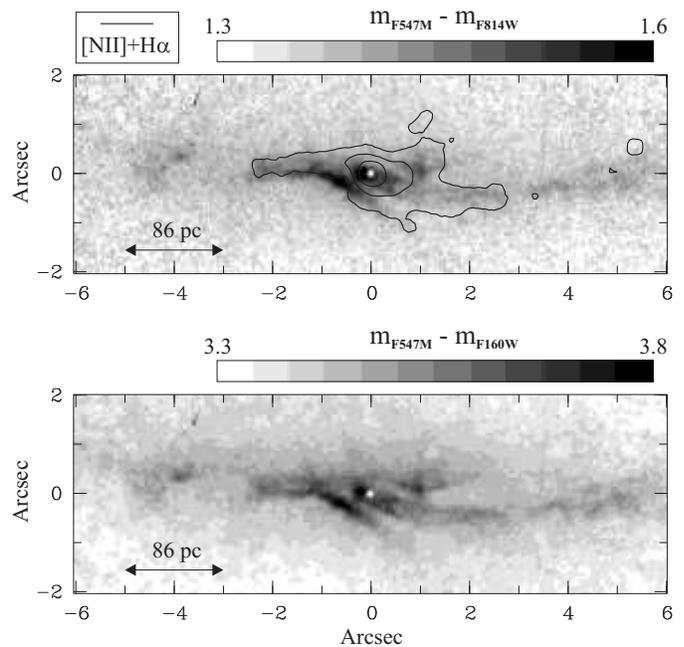}
	 \caption{{\bf Top panel:} $V-I$ colour map, with selected isophotes (5, 20
	 and 100 $\times$ \ten{-18} \Wmarcsec) of the HST \NII+\Ha\ image overimposed.
	 {\bf Bottom panel:} $V-H$ colour map.}
	 \label{FigDust}
      \end{figure}
      
      The colour maps (Fig.~\ref{FigDust}) outline the (patchy) dust structures
      present in the central region of M\,104. As already pointed out in Paper~1,
      there is a straight-line dust lane southeast of the nucleus. It has a projected
      length $> 1\farcs2$ ($\sim 50$~pc). Taking a minimum inclination of 82\degr\ for
      the galaxy and assuming that this dust lane lies in the equatorial plane of the
      galaxy, this yields a deprojected length $> 370$~pc. The associated $E(V-I)$
      reaches 0.2~mag,  corresponding to an apparent A$_V$ $\sim$ 0.5~mag (for $R_V =
      3.1$). Especially in the $V-H$ colour map (Fig.~\ref{FigDust}, bottom panel),
      there is a hint for the existence of a symmetric, though much fainter,
      extinction feature on the northern side of the nucleus.

      Last, the nucleus of the galaxy appears extremely blue (local $V-I$ difference of 
      0.3~mag) compared to its surroundings. It is unlikely that this feature is an
      artifact\footnote{We checked that it is still present after harmonisation of the
      spatial resolutions in the F547M, F814W, and F160W images.},  as already
      mentioned in K+96.
      
\section{Discussion}
   \label{SecDis}
   
   Recently, Regan \& Mulchaey (\cite{Regan99}) argued that nuclear bars may not be
   the primary agent for the fueling of active galactic nuclei (AGN) and suggested
   nuclear spirals as an alternative mechanism. Martini \& Pogge (\cite{MP99})
   also analysed visible and near-infrared HST images of a sample of 24 Seyfert 2
   galaxies and found that 20 of these exhibit nuclear spirals (with only 5 clear
   nuclear bars). Physical processes involved in the formation of such spirals are not
   clear yet, although acoustic instabilities have often been mentioned as a possible 
   mechanism in non self-gravitating nuclear discs (Montenegro \etal~\cite{MYE99}, 
   Elmegreen \etal~ \cite{Elm+98}, and references therein). In M~104 (classified as a
   liner), kiloparsec-scale spiral arms are indeed present but a straight dust lane is
   also found closer to the nucleus: it is a specific signature of strong bars with
   inner Lindblad resonance  (Athanassoula \cite{Lia92}). If it actually traces a
   nuclear bar, this dust lane should have a symmetric  counterpart on the other side
   of the nucleus. There is indeed a hint for such a  feature (see
   Sect.~\ref{SecDust}) but it is rather weak. However, this weakness  could be
   attributed to the fact that this second dust lane would be located on the far
   (north) side of the galaxy. 
   
   To test this hypothesis, we computed the effect of the presence of two symmetric
   dust filaments using the luminosity density model of M~104 of Emsellem
   (\cite{Ems95}, Paper~1 hereafter). We assumed the filaments to be in the equatorial
   plane of the galaxy  (10\degr\
   from end-on), and their characteristics were set to ensure that the  mean extinction
   for the southern filament was consistent with the observations. The model predicts
   that the apparent extinction should rapidly decrease northward, with apparent $A_V
   < 0.03$~mag or $E(V-H) < 0.026$, $0\farcs6$ north of the nucleus (for $R_V =
   3.1$). These are upper limits since dust scattering and clumpiness would tend to
   significantly reduce these values. For these extinction levels, we indeed expect
   the northern filament to be barely detectable in our colour maps.
   
   The observed kinematics would also fit naturally into the strong nuclear bar
   picture. The fact that overall the velocities are small compared to the predicted 
   circular velocities, obviously argue for the presence of strongly non-circular
   motions. The velocity profile in the spiral features could be explained by the
   combination of the streaming motions and projection effects as the spiral curves
   around the nucleus and becomes perpendicular to the line-of-sight. If the nuclear
   bar hypothesis is correct, the existence of offset straight dust lanes
   also requires the presence of an inner Linblad resonance (ILR), and an extended
   $x_2$ orbit family (e.g. Athanassoula \& Bureau \cite{Lia99}, hereafter AthB99). 
   This would explain the observed kinematical decoupling between the gas in the
   nucleus and in the spiral arms, as clearly illustrated in the models of AthB99.
   The observed asymmetry in the central velocity gradients agrees also qualitatively
   with the predictions made by AthB99, when dust is included (see their Fig.~10 with
   e.g. $\psi = 22.5$\degr). We should however keep in mind that significant
   differences exist between the gas distribution, as observed in the core of M\,104,
   and as idealised in the model of AthB99. 
   
   The presence of a large-scale bar has already been suggested in Papers~1 and 2
   (pattern speed $\Omega_p$ of $\sim$120~\kms\ kpc$^{-1}$ for a distance to M\,104 of
   8.8 Mpc). The mass model predicted a strong ILR for this primary bar,
   located roughly 20\arcsec\ from the nucleus. The secondary nuclear bar
   discussed in this letter would then be well inside this resonance. From the
   extension and orientation of the central dust lanes, we can roughly estimate a
   semi-major axis of $a = 425$~pc ($\sim$10\arcsec), and an orientation of
   approximately 10\degr\ from end-on\footnote{See Gerhard \cite{Ger89}} for this bar.
   The location of the corotation can be estimated from the relation $r_L \simeq
   1.2 \times a$ (AthB99, $r_L$ is the Lagrangian radius), yielding a value of
   $\sim$12\arcsec. This is right where the transition region between
   the inner and outer disks occurs (Seifert \& Scorza \cite{Seifert96}). More
   detailed modelling is needed to accurately estimate its pattern speed.
   
   If there are clearly some hints of the presence of a nuclear bar in M\,104, as
   discussed above, it is also clear that additional information are needed to confirm
   or infirm its existence. In particular, the main support for the nuclear bar
   scenario comes from the presence of the straight, southern dust filament, which
   could actually be much further from the nucleus than we assumed (e.g. if it is
   outside the equatorial plane of the galaxy). In the same way, the
   kinematical decoupling of the central regions could alternatively be due to, e.g. a
   nuclear keplerian disk. The answer to this should come soon,  from HST/STIS
   observations of M\,104. These high spatial resolution data will allow detailed
   comparison between the observed PVDs and the gaseous kinematics predicted by
   various models (nuclear bar, keplerian disk). Emission lines in the infrared, where
   dust is less problematic, could also be very valuable to understand the gas
   distribution and kinematics in more details.
	
\begin{acknowledgements}
 PF acknowledges support by the R\'egion Rh\^one-Alpes under an Emergence fellowship.
\end{acknowledgements}


\end{document}